\documentclass[conference]{IEEEtran}
\IEEEoverridecommandlockouts
\usepackage{cite}
\usepackage{amsmath,amssymb,amsfonts}
\usepackage{algorithmic}
\usepackage{graphicx}
\usepackage{textcomp}
\usepackage{xcolor}
\usepackage{threeparttable}
\usepackage{subcaption}
\usepackage{hyperref}
\usepackage{enumitem}
\usepackage{xspace}

\def\BibTeX{{\rm B\kern-.05em{\sc i\kern-.025em b}\kern-.08em
    T\kern-.1667em\lower.7ex\hbox{E}\kern-.125emX}}
\begin{document}

\title{Cross-chain Sharing of Personal Health Records: Heterogeneous and Interoperable Blockchains}
% \title{Secure Cross-chain PHR Exchange Using A Heterogeneous and Interoperable Blockchain}
% \title{\schemeName：A Scheme for Sharing Personal Health Records Across Heterogeneous and Interoperable Blockchains}
% \\

% \author{\IEEEauthorblockN{Yongyang Lv}
% \IEEEauthorblockA{\textit{College of Intelligence and Computing} \\
% \textit{Tianjin University}\\
% Tianjin, China \\
% lvyongyang@tju.edu.cn}
% \and
% \IEEEauthorblockN{Ruitao Feng}
% \IEEEauthorblockA{\textit{SCIS} \\
% \textit{Singapore Management University}\\
% Singapore, Singapore \\
% rtfeng@smu.edu.sg}
% \and
% \IEEEauthorblockN{Yingwenbo Wang}
% \IEEEauthorblockA{\textit{College of Intelligence and Computing} \\
% \textit{Tianjin University}\\
% Tianjin, China \\
% wangyingwenbo@tju.edu.cn}
% \and
% \IEEEauthorblockN{Kui Chen}
% \IEEEauthorblockA{\textit{College of Intelligence and Computing} \\
% \textit{Tianjin University}\\
% Tianjin, China \\
% chen\_kui@tju.edu.cn}
% \and
% \IEEEauthorblockN{Zhe Hou}
% \IEEEauthorblockA{\textit{School of ICT} \\
% \textit{Griffith University}\\
% Brisbane, Australia \\
% z.hou@griffith.edu.au}
% \and
% \IEEEauthorblockN{Xiaohong Li\IEEEauthorrefmark{*}\thanks{*Corresponding author.}}
% \IEEEauthorblockA{\textit{College of Intelligence and Computing} \\
% \textit{Tianjin University}\\
% Tianjin, China \\
% Xiaohongli@tju.edu.cn}
% }
\DeclareRobustCommand*{\IEEEauthorrefmark}[1]{%
\raisebox{0pt}[0pt][0pt]{\textsuperscript{\footnotesize\ensuremath{#1}}}}

\author{\IEEEauthorblockN{Yongyang Lv\IEEEauthorrefmark{1}, Xiaohong Li\IEEEauthorrefmark{1}, Yingwenbo Wang\IEEEauthorrefmark{1}, Kui Chen\IEEEauthorrefmark{1}, Zhe Hou\IEEEauthorrefmark{2} and Ruitao Feng\IEEEauthorrefmark{3}\IEEEauthorrefmark{,}\IEEEauthorrefmark{*}}
\IEEEauthorblockA{\IEEEauthorrefmark{1}College of Intelligence and Computing, 
Tianjin University, Tianjin, China}
\IEEEauthorblockA{\IEEEauthorrefmark{2}School of Information and Communication Technology, Griffith University, Brisbane, Australia}
\IEEEauthorblockA{\IEEEauthorrefmark{3}Faculty of Science and Engineering, Southern Cross University, Gold Coast, Australia}
\IEEEauthorblockA{\IEEEauthorrefmark{*}Corresponding author}
\IEEEauthorblockA{Emails: lvyongyang@tju.edu.cn,  Xiaohongli@tju.edu.cn, wangyingwenbo@tju.edu.cn,\\chen\_kui@tju.edu.cn, z.hou@griffith.edu.au, ruitao.feng@scu.edu.au}
}

\maketitle

\begin{abstract}
With the widespread adoption of medical informatics, a wealth of valuable personal health records (PHR) has been generated. Concurrently, blockchain technology has enhanced the security of medical institutions. However, these institutions often function as isolated data silos, limiting the potential value of PHRs. As the demand for data sharing between hospitals on different blockchains grows, addressing the challenge of cross-chain data sharing becomes crucial. When sharing PHRs across blockchains, the limited storage and computational capabilities of medical Internet of Things (IoT) devices complicate the storage of large volumes of PHRs and the handling of complex calculations. Additionally, varying blockchain cryptosystems and the risk of internal attacks further complicate the cross-chain sharing of PHRs. This paper proposes a scheme for sharing PHRs across heterogeneous and interoperable blockchains. Medical IoT devices can encrypt and store real-time PHRs in an InterPlanetary File System, requiring only simple operations for data sharing. An enhanced proxy re-encryption(PRE) algorithm addresses the differences in blockchain cryptosystems. Multi-dimensional analysis demonstrates that this scheme offers robust security and excellent performance.
\end{abstract}


\section{Introduction}\label{I}
The widespread adoption of medical informatization has led to the creation of numerous Personal Health Records (PHR), these records contain sensitive physiological data parameters and patient medical histories, emphasizing high privacy concerns\cite{b1}. However, due to insecure current sharing mechanisms and unclear data rights and responsibilities, medical institutions that collect PHRs often operate as isolated data silos, limiting the full potential utilization of PHRs' value \cite{b2}. As blockchain technology gains popularity, more medical institutions are leveraging it to enhance data-sharing capabilities, aiming to address issues such as single points of failure and trust between different institutions during data sharing \cite{b2}\cite{b7}\cite{b8}\cite{b9}\cite{b10}. However, the above solutions primarily focus on data sharing within the same blockchain network, neglecting scenarios where medical institutions operate on different blockchains. 
%We illustrate the process of cross-chain PHR sharing with an example. Hospital A in Blockchain A collects various PHR in real-time through IoT devices such as wearable devices. However, due to the limited storage and computational capabilities of these devices, they encrypt the PHR with their private key and store it in InterPlanetary File System (IPFS) \cite{b1}\cite{b10}. When Hospital B in Blockchain B needs to access the data from one of Hospital A's IoT devices, the issue of cross-chain PHR sharing arises. Compared to other PHR sharing scenarios, cross-chain sharing presents several key challenges:

We illustrate the real need for cross-chain sharing of PHRs with an example. When Alice was hospitalized at Hospital A (on blockchain A), her PHRs were continuously collected via the hospital A’s medical IoT devices. Due to the limited storage and computational capabilities of this equipment, Alice's PHRs were encrypted using her private key and sent to Hospital A's InterPlanetary File System (IPFS) for storage\cite{b1}\cite{b10}. Later, when Alice seeks treatment at Hospital B, Doctor Bob needs access to her previous PHRs from Hospital A. Since Hospitals A and B are on different blockchains, this scenario requires cross-chain PHR sharing. The main challenges in sharing PHRs across chains are: 1) Different blockchains typically employ distinct encryption mechanisms. 2) PHR is encrypted and stored on IPFS, with cross-chain sharing of PHR necessitating complex procedures \cite{b2}\cite{b7}\cite{b10}. 3) The cross-chain data sharing faces external attacks \cite{b11}\cite{b12}.

%To address the complexity and inefficiency of traditional data sharing methods, Proxy Re-Encryption (PRE) algorithms have been widely applied in existing data sharing schemes \cite{b13}\cite{b14}\cite{b15}. The advantage of PRE is that a semi-trusted proxy can convert a ciphertext generated by the data owner into a form that can be decrypted by the data user using their private key. This method eliminates the need for data downloading, decryption, and re-encryption, thereby simplifying the data sharing process. In this simplified process, the data owner only needs to generate a re-encryption key, which the proxy uses to transform the ciphertext without accessing the data owner's private key. However, in practice, different medical institutions may use different cryptographic systems, and existing PRE algorithms are generally based on the assumption of a single cryptographic mechanism, without addressing internal attack issues \cite{b6}. More importantly, existing research \cite{b6}\cite{b7}\cite{b8}\cite{b2}\cite{b9} has not adequately solved the problem of PHR sharing between blockchains with different cryptographic systems (termed as cross-heterogeneous chain data sharing in this paper).
Existing research often treats blockchain as a trusted entity for achieving medical data sharing through various encryption technologies. Wang et al. \cite{b8} proposed a decentralized electronic medical record-sharing framework called MedShare, which designed a constant-size attribute-based encryption (ABE) scheme to achieve fine-grained access control. Quan et al. \cite{b2} proposed a reliable medical data-sharing framework in an edge computing environment, addressing the challenges of real-time, multi-attribute authorization in ABE through a blockchain-based distributed attribute authorization strategy (DAA). Banik et al. \cite{b9} utilized public key encryption with keyword search (PEKS) technology to design a federated blockchain with preselected users, achieving data security, access control, privacy protection, and secure search. Liu et al. \cite{b7} combined ABE and searchable encryption (SE) to propose a multi-keyword search-based data-sharing scheme, providing comprehensive privacy protection and efficient ciphertext retrieval for electronic medical records. Zhao et al. \cite{b23} proposed a large-scale, verifiable and privacy-preserving dynamic fine-grained access control scheme based on attribute-based proxy re-encryption (PRE). The PRE algorithm is widely used in existing data sharing schemes\cite{b5}\cite{b11}.  However, the cryptographic systems among different medical institutions can vary significantly. The encryption algorithms in \cite{b8}\cite{b2}\cite{b9}\cite{b7}\cite{b23}\cite{b5}\cite{b11} assume uniform cryptographic mechanisms, making them unsuitable for real medical scenarios. 
%Additionally, these schemes are not designed for medical IoT devices with low storage and computational capabilities, nor do they address potential internal attacks during data sharing\cite{b4}\cite{b11}\cite{b21}.

To address these challenges, this paper enhances the proxy re-encryption algorithm from \cite{b12}, enabling PHRs ciphertext to be converted and decrypted between Identity-Based Encryption (IBE) and Certificateless Cryptography (CLC) systems. Based on this improvement, we develop a cross-chain sharing scheme for PHR. In this scheme, real-time generated PHRs are encrypted and stored in the IPFS. When data sharing is required, IoT terminal devices with limited storage and computational capabilities can facilitate PHR sharing by utilizing smart contracts.  The main contributions of this paper are as follows:
%Additionally, we integrate a Cryptographic Reverse Firewall (CRF) and a blockchain audit mechanism into the scheme, ensuring protection against internal attacks and preventing information leakage.
\begin{enumerate}[leftmargin=*]
\item We introduce an enhanced proxy re-encryption algorithm capable of facilitating data sharing between IBE and CLC. 
\item Building upon the enhanced proxy re-encryption algorithm, we present a scheme for cross-chain PHR sharing. 
\item Security and performance evaluations demonstrate that the proposed scheme not only meets stringent security criteria but also exhibits superior operational efficiency.
    % \item The scheme is proposed for sharing PHRs across heterogeneous and interoperable blockchains, facilitating PHR sharing using IBE and CLC cryptosystems.
    % \item The scheme enables medical IoT devices with limited performance to manage PHR sharing through straightforward operations.
    % \item Analytical results from security, and performance tests demonstrate that this scheme meets stringent security requirements and delivers superior performance.
\end{enumerate}

\section{Preliminaries}\label{P}
This section introduces the concepts of bilinear pairings, which are essential for constructing the scheme described in Section \ref{C}.

\subsection{Bilinear Pairing}

Let $G_1$ and $G_2$ be two multiplication groups of order prime $q$, with $g$ as the generator of $G_1$. A bilinear pairing $e:G_1 \times G_1\rightarrow G_2$ satisfies the following properties:

\begin{enumerate}[leftmargin=*]
\item Bilinearity: For $\forall\left(g_1,g_2\right)\in G_1$, $\forall\left(a,b\right)\in Z_q^\ast$, it must hold that $e\left(g_1^a,g_2^b\right)=e\left(g_1,g_2\right)^{ab}$.
\item Non-degeneracy: For $\exists\left(g_1,g_2\right)\in G_1$ and $1_{G_2}$ be the identity element of $G_2$, there have $e\left(g_1,g_2\right)\neq1_{G_2}$.
\item Computability: For $\forall\left(g_1,g_2\right)\in G_1$, there exists an effective algorithm to compute $e\left(g_1,g_2\right)$.
\end{enumerate}

% \subsection{Cryptographic Reverse Firewalls}
% In \cite{b19}, let $W$ be a CRF, $P=\left(receive,next,output\right)$ be a party, we can say $W$ is a CRF for $P$ if it meets the following properties. Here, $\sigma$ is an initial public parameter, $m$ is the transmitted message. Define $W\circ P$ as follows:
% \vspace{-1ex}
% \begin{gather*}
%     W\circ P {:=}(receive_{W\circ P}\left(\sigma,m\right)=receive_P\left(\sigma,W\left(m\right)\right)\\
%     next_{W\circ P}\left(\sigma\right) =W\left(next_P\left(\sigma\right)\right)\\
%     output_{W\circ P}\left(\sigma\right) =output_P\left(\sigma\right))
% \end{gather*}
% A qualified CRF needs to satisfy the following properties:

% (1) Maintain functionality. If the user’s computer operates correctly, the CRF will not compromise the functionality of the cryptographic algorithms.

% (2) Preserve security. Regardless of how the user’s computer is affected by an attacker, the use of the CRF will remain as secure as the correct execution of cryptographic algorithms.

% (3) Resist the exfiltration. No matter how to run the user’s computer, the CRF will prevent the computer from leaking confidential information.

\section{CONSTRUCTION}\label{SO}
This section provides a detailed explanation of the system model and the scheme's implementation. 

\begin{figure}[htbp]
    \centering
    \includegraphics[width=8cm]{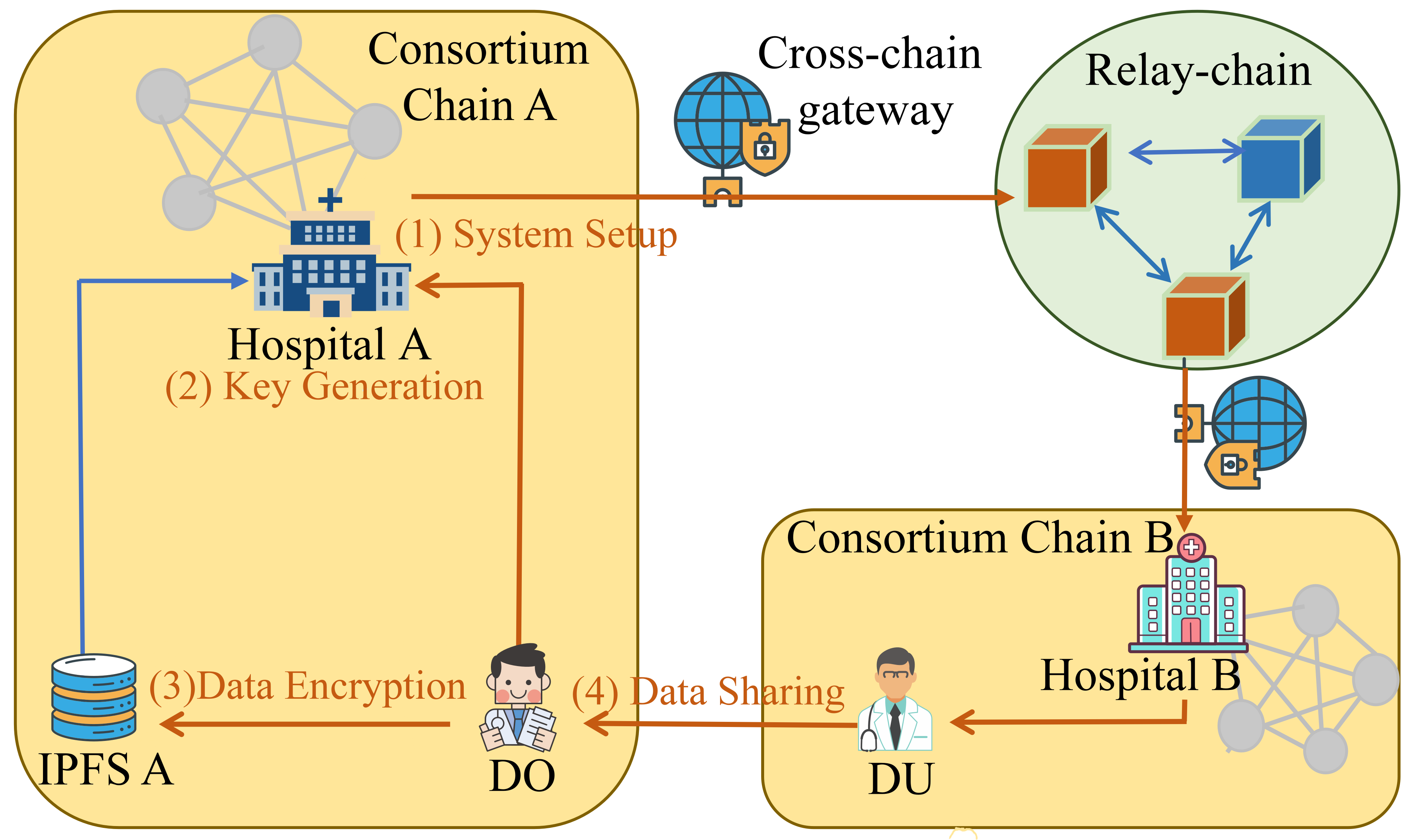}
    \caption{The Scheme Model}
    \label{structure}
\end{figure}

\subsection{System Model}
In the scheme, we assume that there is a node, ${\text{Hospital}_\text{A}}$, in consortium chain A, which uses IBE as its cryptographic system. Similarly, there is a node, ${\text{Hospital}_\text{B}}$, in consortium chain B, which uses CLC as its cryptographic system. This scheme assumes that a data user in ${\text{Hospital}_\text{B}}$ needs to access some PHRs from a data owner in ${\text{Hospital}_\text{A}}$. The scheme model is shown in \textbf{Fig. \ref{structure}}, and the scheme includes the following entities:
%The process of sharing PHR is described in detail in this paper. 

\textbf{Hospital (${\text{Hospital}_\text{i}}$)}: A node in the blockchain that generates keys for users within the chain.

\textbf{Data Owner (DO)}: A user in ${\text{Hospital}_\text{A}}$ who owns the PHR.

\textbf{Data User (DU)}: A user in ${\text{Hospital}_\text{B}}$ who can be a patient, doctor, researcher, or any other person needing to use PHR.

\textbf{Relay Chain}: It provides computing power and is responsible for the calculation of re-encrypted ciphertext. 

\textbf{Interplanetary File System (${\text{IPFS}_\text{i}}$)}: A semi-trusted distributed database responsible for storing PHRs to reduce the storage burden on IoT terminal devices.

%\textbf{Firewall (${\text{CRF}_\text{i}}$)}: Deployed between user terminal equipment and blockchain nodes.

\subsection{Scheme Construction}\label{C}

This section introduces the details of the scheme's implementation.\par

\subsubsection{System Setup} At this stage, consortium chains A and B register their systems to generate system parameters. 

\begin{enumerate}
\item [(1.1)] Given a security parameter $k$, the ${\text{Hospital}_\text{A}}$ in consortium chain A selects $s \in Z_q^*$ as the master private key and calculates the system public key $h_1=g^s$, where $g$ is the generator of $G_1$. ${\text{Hospital}_\text{A}}$ public parameters $par_1=\{G_1,G_2,e,g,h_1,H_1,H_2\}$.
\item [(1.2)] Similarly, ${\text{Hospital}_\text{B}}$ in consortium chain B randomly selects $y\in Z_q^*$ as the master private key and calculates the system public key $h_2=g^y$, public the parameters to $par_2=\{G_1,G_2,e,g,h_2,H_1,H_2\}$.
\end{enumerate}

\subsubsection{Key Generation} 
At this stage, consortium chains A and B generate keys for users in their respective chains.
\begin{enumerate}
    \item [(2.1)] ${\text{Hospital}_\text{A}}$ generates the user's public key $pk_{DO}=H_1\left(ID_{DO}\right)$ and private key $sk_{DO}=pk_{DO}^s$, and sends $sk_{DO}$ to the DO.
    
    \item [(2.2)] ${\text{Hospital}_\text{B}}$ generates the user's partial private key $D_{DU}=H_1\left(ID_{DU}\right)^y$. ${\text{Hospital}_\text{B}}$ sends $D_{DU}$ to the DU. The DU randomly selects $r\in Z_p^\ast$, calculates the private key $sk_{DU}=(D_{DU}{)}^r=H_1\left(ID_{DU}\right)^{yr}$, and the public key $pk_{DU}=(pk_{DU1},pk_{DU2})=\left(H_1\left(ID_{DU}\right),\ (h_2{)}^r\right)$.
\end{enumerate}

\subsubsection{Data Encryption}
At this stage, the DO encrypts PHR and uploads it to ${\text{IPFS}_\text{A}}$ for storage.

\begin{enumerate}
    \item [(3.1)] The DO selects the message $M$ (containing PHR) to be shared, given $par_1$ and $pk_{DO}$, randomly selects $\alpha\in Z_q^\ast$, and generates the ciphertext  $C_{DO}=(c_1,c_2)$,  $c_1=g^\alpha, c_2=M\cdot e(h_1,pk_{DO})^\alpha$. Then, the DO sends $C_{DO}$ and its identifier $Data_1$ to ${\text{IPFS}_\text{A}}$ for storage.
    \item [(3.2)] Simultaneously, ${\text{Hospital}_\text{A}}$ saves the ciphertext identifier $Data_1$ and its address $Add_1$ in the access list $List_1$.
    % \begin{gather*}
    %      c_1\prime=c_1 \cdot g^\beta \quad
    %     c_2\prime=c_2 \cdot e(h_1',pk_{DO})^\beta \quad c_3\prime=pk_{DO}^\beta
    % \end{gather*}
\end{enumerate}

\subsubsection{Data Sharing}
At this stage, the DU initiates a cross-chain access request to the DO. After successfully verifying the request message, the DO shares the data with the DU.

\begin{enumerate}
    \item [(4.1)] To access the message $M$ from the DO, the DU must first send a cross-chain access request message $M_1 = \{request_1,pk_{DO},pk_{DU},T_1,N_1\}_{pk_{DO}}$. Here, $request_1$ is the cross-chain access identifier, $T_1$ is the timestamp, and $N_1$ is the nonce to maintain session freshness. The message $M_1$ is forwarded to the DO via the cross-chain gateway.
    \item [(4.2)] Upon receiving the request, the DO verifies the validity of the message and the correctness of the DU's identity. If the verification is successful, the DO randomly selects $\lambda$ and $X$. Then, using its own private key $sk_{DO}$ and the DU's public key $pk_{DU}$, the DO generates the re-encryption key $rk_{DO} = ( {rk}_1, {rk}_2, {rk}_3 )$, $rk_1 = H_2(X) / sk_{DO}, rk_2 = g^\lambda,
          rk_3 = X \cdot e(pk_{DU1}, pk_{DU2})^\lambda$, and sends it to ${\text{Hospital}_\text{A}}$.
    % \begin{gather*}
    %      rk_1 = H_2(X) / sk_{DO}' \quad rk_2 = g^\lambda \\
    %      rk_3 = X \cdot e(pk_{DU1}, pk_{DU2})^\lambda
    % \end{gather*}
%    \item [(4.3)] Upon receiving $rk_{DO}$, ${\text{CRF}_\text{A}}$ generates the randomized re-encryption key $rk_{DO}' = (rk_1', rk_2', rk_3')$, $\quad rk_1' = rk_1 \cdot pk_{DO}^{-\beta}, rk_2' = rk_2 \cdot g^\beta, rk_3' = rk_3 \cdot e(pk_{DU1}, pk_{DU2})^\beta$.
    % \begin{gather*}
    %      \quad rk_1' = rk_1 \cdot pk_{DO}^{-\beta} \quad rk_2' = rk_2 \cdot g^\beta \\ rk_3' = rk_3 \cdot e(pk_{DU1}, pk_{DU2})^\beta
    % \end{gather*}
    \item [(4.3)] The DO then sends the ciphertext identifier $Data_1$ and the cross-chain data sharing permission message $M_2=\{request_2,pk_{DO},pk_{DU},rk_{DO},T_2,N_2\}$ to ${\text{Hospital}_\text{A}}$. ${\text{Hospital}_\text{A}}$ ultimately sends $M_2$ and the ciphertext $C_{DO}$ to the relay chain.
    \item [(4.4)] The relay chain generates the re-encrypted ciphertext $C_{DU} = (C_1, C_2, C_3, C_4)$ based on the given $C_{DO}$ and $rk_{DO}$, $C_1 = c_1, C_2 = c_2 \cdot e(C_1, rk_1), 
        C_3 = rk_2, C_4 = rk_3$. Finally, the relay chain sends the response message $M_3=\{respond_1,C_{DU},T_3,N_3\}$ to the DU via the cross-chain gateway.
    %The relay chain, upon receiving the permission information, authenticates the validity of the request through the coordination node. If the request authentication is successful, 
    % \begin{gather*}
    %     C_1 = c_1' \quad C_2 = c_2' \cdot e(C_1, rk_1' \cdot c_3')\\
    %     C_3 = rk_2' \quad C_4 = rk_3'
    % \end{gather*}
    \item [(4.5)] Upon receiving the response message $M_3$, the DU first verifies the validity of the message. After successful verification, the DU uses its private key $sk_{DU}$ to calculate $X = C_4 / e(C_3, sk_{DU})$, and then calculates $M = C_2 / e(C_1, H_2(X))$ to obtain the message $M$.
\end{enumerate}

\subsection{Scheme Correctness Proof}
We check whether the DU has accurately decrypted the re-encrypted ciphertext $C_{DU} = (C_1, C_2, C_3, C_4)$.

\vspace{-1ex}
\begin{align*}
    \frac{C_4}{e\left(C_3,sk_{DU}\right)}&=\frac{rk_3}{e\left(rk_2,sk_{DU}\right)}\\
    &=X \cdot \frac{e(pk_{DU1},pk_{DU2}) ^{\lambda}}{e(g^{\lambda},sk_{DU})}\\
    &=X \cdot \frac{e(H_1\left(ID_{DU}\right),g^{yr}) ^{\lambda}}{e(g^{\lambda},H_1\left(ID_{DU}\right)^{yr})}\\
    &=X
\end{align*}

It is evident that $X$ can be correctly decrypted by DU.
\vspace{-1ex}
\begin{align*}
    \frac{C_2}{e\left(C_1,H_2\left(X\right)\right)}&=\frac{c_2\cdot e(c_1,rk_1)}{e(c_1,H_2(X))}\\
    &=M \cdot \frac{e\left(g^{s},pk_{DO}\right)^{\alpha}\cdot e(g^{\alpha},H_2(X))}{e\left(g^{\alpha},H_2\left(X\right)\right) \cdot e(g^{\alpha},sk_{DO})}\\
    &=M \cdot \frac{e\left(g^{s},pk_{DO}\right)^{\alpha}}{e\left(g^{\alpha},pk_{DO}^{s}\right)}\\
    &=M
\end{align*}

Based on the correct decryption of $X$, the DU also correctly decrypts the ciphertext $C_{DU}$ to obtain message $M$.
% \begin{table}[h!]
% \centering
% \begin{tabular}{p{8cm}}
% \hline
% \textbf{${\text{Algorithm}_\text{1}}$}\\
% \hline
% \textbf{Input: }$Data_1$, $M_i$(request, $ID_{DU}$, Timestamp) \\
% \textbf{Output: }True/False\\
% \textbf{Begin:} \\
% \quad 1. ${\text{Hospital}_\text{i}}$ receives message $(Data_1, M_i)$\\
% \quad 2. // get the number of DU's access from access list\\
% \quad 3. accessCount $\gets$ countOf($List_1$, find($ID_{DU}$))\\
% \quad 4. \textbf{if !}accessCount $<$ maxAccessCount \textbf{then}\\
% \quad 5. \quad Display ``Access limit reached!''\\
% \quad 6. \quad \textbf{return} False\\
% \quad 7. // search the address of ciphertexts marked by $Data_1$ in the blockchain\\
% \quad 8. $addr_1 \gets$ searchInChain($Data_1$)\\
% \quad 9. \textbf{if !}$addr_1$ exists \textbf{then}\\
% \quad 10. \quad Display "Target data doesn’t exist"\\
% \quad 11. \quad \textbf{return} False\\
% \quad 12. // get ciphertext in IPFS based on address $addr_1$\\
% \quad 13. $C_{DO}' \gets$ searchInIPFS$_\text{1}$($addr_1$)\\
% \quad 14. ${\text{Hospital}_\text{1}}$ sends message $(C_{DO}', M_i)$ to RelayChain\\
% \quad 15. // record this access\\
% \quad 16. $List$.insert($ID_{DU}$, Timestamp, $Data_1$)\\
% \quad 17. \textbf{return} True\\
% \textbf{end}\\
% \hline
% \end{tabular}
% \end{table}

\section{Safety Analysis}\label{CSA}
% This section first proves the correctness of the \schemeName scheme. Secondly, we use Ban logic and CPA security analysis to prove the security of the \schemeName scheme.

% \subsection{Scheme Correctness Proof}

% \textbf{Definition:} If DO encrypts message $M$ to generate ciphertext $CT$, and the re-encrypted ciphertext is $CT’$, then proxy re-encryption algorithm is correct if $Decrypt_{DU}(CT’) = M$.

% We verify the correctness of the \schemeName scheme by checking if the DU can accurately decrypt the re-encrypted ciphertext $C_{DU} = (C_1, C_2, C_3, C_4)$.

% \vspace{-1ex}
% \begin{align*}
%     \frac{C_4}{e\left(C_3,sk_{DU}\right)}&=\frac{rk_3\prime}{e\left(rk_2\prime,sk_{DU}\right)}\\
%     &=X \cdot \frac{e(pk_{DU1},pk_{DU2}) ^{\lambda+\beta}}{e(g^{\lambda+\beta},sk_{DU})}\\
%     &=X \cdot \frac{e(H_1\left(ID_{DU}\right),g^{ybr}) ^{\lambda+\beta}}{e(g^{\lambda+\beta},H_1\left(ID_{DU}\right)^{ybr})}\\
%     &=X
% \end{align*}

% It is evident that $X$ can be correctly decrypted by DU.
% \vspace{-1ex}
% \begin{align*}
%     \frac{C_2}{e\left(C_1,H_2\left(X\right)\right)}&=\frac{c_2'\cdot e(c_1',rk_1'\cdot c_3')}{e(c_1',H_2(X))}\\
%     &=M \cdot \frac{e\left(g^{sa},pk_{DO}\right)^{\alpha+\beta}\cdot e(g^{\alpha+\beta},H_2(X))}{e\left(g^{\alpha+\beta},H_2\left(X\right)\right) \cdot e(g^{\alpha+\beta},sk_{DO}')}\\
%     &=M \cdot \frac{e\left(g^{sa},pk_{DO}\right)^{\alpha+\beta}}{e\left(g^{\alpha+\beta},sk_{DO}\prime\right)}\\
%     &=M
% \end{align*}

% Based on the correct decryption of $X$, the DU also correctly decrypts the ciphertext $C_{DU}$ to obtain message $M$.

In this section, we prove that the scheme meets CPA security. We references the security model played by the  challenger $\mathcal{C}$ and the adversary $\mathcal{A}$ in \cite{b12}, which is detailed as follows:

\textbf{Theorem 1.} The KGC of the data owner in this scheme preserve weak security and resist weak exfiltration. Here, preserving weak security means that Ateniese's \cite{b20} IBPRE scheme is CPA secure. Resisting weak exfiltration means that when faced with an adversary launching an ASA that does not affect normal functionality, this scheme can resist information leakage.

\textbf{Proof.} Similar to the proof method in \cite{b12}, our constructed scheme also achieves CPA security.

\section{Performance Analysis}\label{PA}
%%%%%%%%%%%%%%%%%%%%%%%%%%%%%%%%% Performance %%%%%%%%%%%%%%%%%%%%%%%%%%%%%%%%%%%%%
%%%%%%%%%%%%%%%%%%%%%%%%%%%%%%%%%% Analysis %%%%%%%%%%%%%%%%%%%%%%%%%%%%%%%%%%%%%%%
In this section, we evaluate the computational and communication overhead. Experiments were conducted on a Lenovo laptop with an AMD Ryzen 7 5800H processor with Radeon Graphics running at 3.20 GHz and 16GB of RAM. The host machine runs Ubuntu 22.04.2 operating system, with Java 1.8.0\_102, and employs FISCO-BCOS blockchain v3.6.0. 
\begin{table*}[htbp]
\centering
\renewcommand{\arraystretch}{1.3}
\begin{threeparttable}[c]
\caption{Comparison of Communicational Overhead}
\label{table:communicationalOverhead}
\begin{tabular}{|l|c|c|c|c|c|c|}
\hline
\bfseries Scheme & \bfseries Key$_{DO}$ & \bfseries Key$_{DU}$ & \bfseries CT & \bfseries RK & \bfseries CT’ & \bfseries Total(bytes)\\
\hline
\bfseries CP-HAPRE\cite{b18} & (2n+4)$|G_1|$ & (2n+4)$|G_1|$ & (3N+2)$|G_1|$+$|G_2|$ & 7$|G_1|$ & 4$|G_1|$+$|G_2|$ & 4864\\
\hline
\bfseries CDSS\cite{b21} & 6$|G_1|$+2$|Z_q|$ & 7$|G_1|$+2$|Z_q|$ & 3$|G_1|$+2$|G_2|$ & 4$|G_1|$+$|G_2|$ & $|G_1|$+3$|G_2|$ & 2408\\
\hline
\bfseries ABE-IBE\cite{b17} & (2N+1)$|G_1|$ & 2$|G_1|$ & (N+1)$|G_1|$+$|G_2|$ & (4N+3)$|G_1|$+$|G_2|$ & 2$|G_1|$+$|G_2|$ & 4928\\
\hline
\bfseries Ours & 2$|G_1|$ & 3$|G_1|$ & 2$|G_1|$+$|G_2|$ & 2$|G_1|$+$|G_2|$ & 2$|G_1|$+2$|G_2|$ & 1344\\
\hline
\end{tabular}
\begin{tablenotes}
\centering
\item $|G_1|$: Storage overhead of group elements in $G_1$(128bytes)\quad$|G_2|$: Storage overhead of group elements in $G_2$(128bytes)\quad \\$|Z_q|$: Storage overhead of group elements in $Z_q$(20bytes)\\

\end{tablenotes}
\end{threeparttable}
\end{table*}
\subsection{Computational Overhead}
We calculate the computational overhead of core operations in the schemes, focusing on computationally expensive operations such as bilinear pairing, exponentiations in the group $G_1$ and $G_2$, and hash function computations. We chose to compare our scheme with \cite{b17}\cite{b18} and \cite{b21} because they, like ours, achieve encryption system transformation through proxy re-encryption. Notably, \cite{b17} first introduced the concept of encryption system transformation. 

\begin{figure}[!t]
    \centering
    \includegraphics[width=9cm]{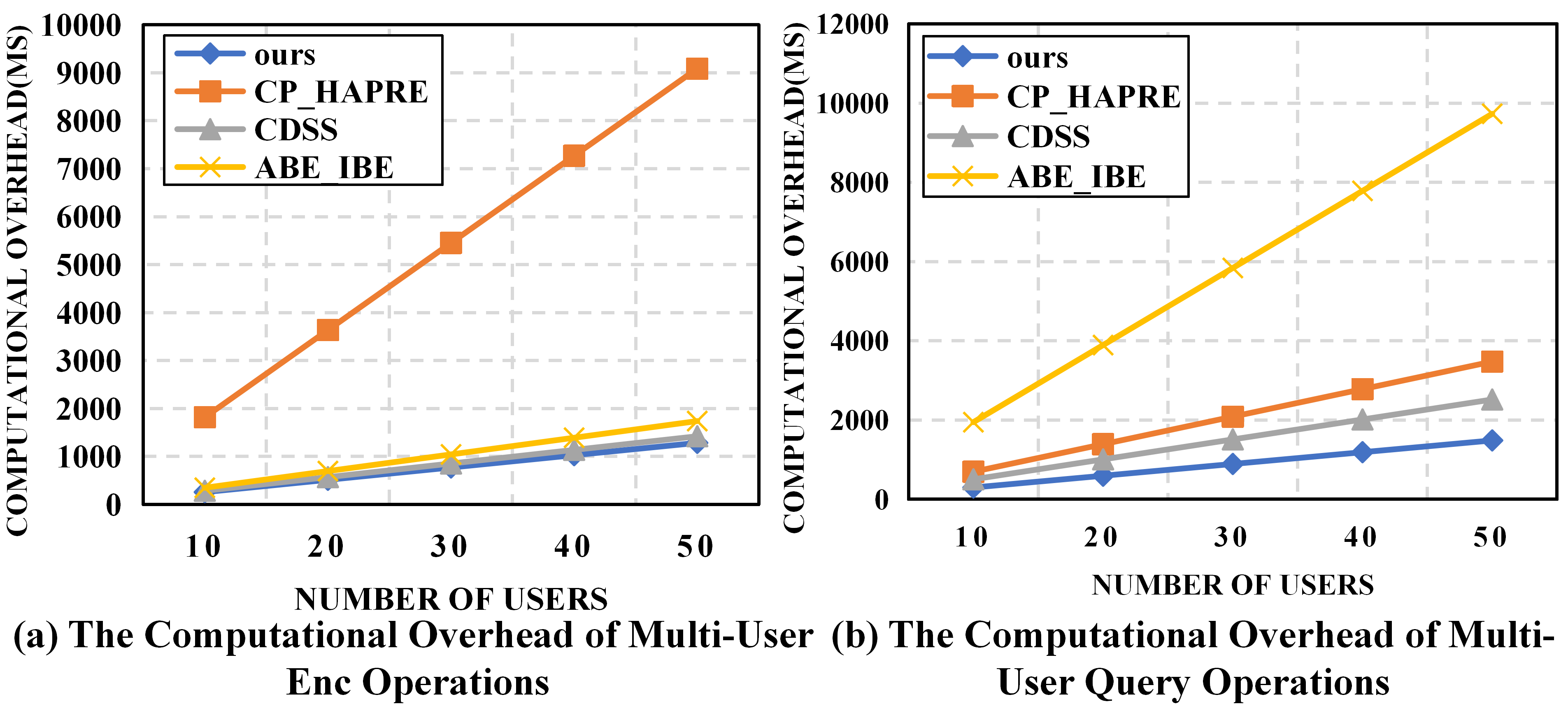}
    \caption{Comparison of Computational Overhead}
    \label{fig:merge}
\end{figure}

As shown in \textbf{Fig. \ref{fig:merge}}(a), as the number of users increases, our scheme exhibits a greater advantage in the time required to complete all users’ \textbf{Enc} operations. As depicted in \textbf{Fig. \ref{fig:merge}}(b), we combine the \textbf{ReKeyGen} and \textbf{ReEnc} operations into a single \textbf{query} operation, the advantage in the time required to complete all \textbf{query} operations becomes more significant as the number of users increases. Thus, compared with \cite{b17}\cite{b18}\cite{b21}, our scheme demonstrates more efficient performance under higher user loads.

\subsection{Communicational Overhead}
Regarding communication overhead, during system operation, it is necessary to transmit and receive key pairs (\textbf{Key}), ciphertexts (\textbf{CT}), re-encryption keys (\textbf{RK}), re-encrypted ciphertexts (\textbf{CT'}), and other primary data. The lengths of these data are directly related to the size of communication overhead during system operation. Therefore, 
We calculate the number of elements contained in the $G_1$, $G_2$ and $Z_q$ group, which are part of the core data of the scheme, to evaluate their communication overhead. It is important to note that the CP-HAPRE and ABE-IBE schemes involve ABE. For fairness in experiments, we use the simplest access policy in ABE, setting the total number of attributes (\textbf{N}) in the access policy to 5 and the number of attributes required for access (\textbf{n}) to 3.

As shown in \textbf{Table \ref{table:communicationalOverhead}}, our scheme enhances system security while maintaining an advantage in communication overhead. Regarding the key \textbf{Key}, on average, each key pair in our scheme only contains 2.5 elements from the $G_1$ group. For the ciphertext \textbf{CT}, our scheme requires only 2 elements from the $G_1$ group and 1 element from the $G_2$ group for ciphertext generation, which contrasts favorably with other schemes. Concerning the re-encryption key \textbf{ReKey}, our scheme needs just 2 elements from the $G_1$ group and 1 element from the $G_2$ group for key generation. As for the re-encrypted ciphertext \textbf{CT'}, communication overhead is generally consistent across all schemes. Therefore, compared with \cite{b17}\cite{b18}\cite{b21}, our scheme can effectively save communication overhead when the number of users increases.

\section{Conclusion}\label{CON}
This paper proposes a cross-heterogeneous chain data sharing scheme tailored for the medical context. It addresses the limitations of storage and computing resources in medical IoT devices and facilitates PHRs sharing through IBE and CLC. The proposed scheme enhances security measures effectively. In the future, we plan to undertake more in-depth research on internal attacks within cross-chain environments to enhance the security of cross-chain protocols. Additionally, we aim to optimize the performance of cross-chain schemes to ensure compatibility with resource-constrained devices, such as those in the internet of medical things.

\section*{Acknowledgment}
This work was supported by the National Key Research and Development Program of China (2021YFF1201102).

% \section*{References}
% Please number citations consecutively within brackets \cite{b1}. The 
% sentence punctuation follows the bracket \cite{b2}. Refer simply to the reference 
% number, as in \cite{b3}---do not use ``Ref. \cite{b3}'' or ``reference \cite{b3}'' except at 
% the beginning of a sentence: ``Reference \cite{b3} was the first $\ldots$''

% Number footnotes separately in superscripts. Place the actual footnote at 
% the bottom of the column in which it was cited. Do not put footnotes in the 
% abstract or reference list. Use letters for table footnotes.

% Unless there are six authors or more give all authors' names; do not use 
% ``et al.''. Papers that have not been published, even if they have been 
% submitted for publication, should be cited as ``unpublished'' \cite{b4}. Papers 
% that have been accepted for publication should be cited as ``in press'' \cite{b5}. 
% Capitalize only the first word in a paper title, except for proper nouns and 
% element symbols.

% For papers published in translation journals, please give the English 
% citation first, followed by the original foreign-language citation \cite{b6}.

% \printbibliography

\end{document}